\documentstyle[12pt]{article}
\begin{document}

\vspace*{-14mm}



\thispagestyle{empty}

\vspace*{35mm}

\begin{center}

{\LARGE \bf On the issue of imposing boundary conditions on quantum fields}

\vspace{5mm}

\medskip

{\sc E. Elizalde}\footnote{Presently on leave at Department of Mathematics,
Massachusetts Institute of Technology, 77 Massachusetts Ave, Cambridge,
MA 02139.
E-mail: 
elizalde@math.mit.edu \
elizalde@ieec.fcr.es} \\
Consejo Superior de Investigaciones Cient\'{\i}ficas (ICE/CSIC) \\
 Institut d'Estudis
Espacials de Catalunya (IEEC) \\ Edifici Nexus, Gran Capit\`{a}
2-4, 08034 Barcelona, Spain \\

\vspace{35mm}

{\small \sl

\hfill As far as the laws of mathematics refer to

\hfill reality, they are not certain; and as far as

\hfill they are certain, they do not refer to reality.

\hfill A. Einstein}

\vspace{30mm}

{\bf Abstract}

\end{center}

An interesting example of the deep interrelation between Physics and
Mathematics is obtained when trying to impose mathematical 
boundary conditions on physical quantum fields. This procedure
has recently been re-examined with  care. Comments on that and 
previous analysis are here provided, together with considerations 
on the results of the purely mathematical zeta-function method, in 
an attempt at clarifying the issue. Hadamard regularization is invoked 
in order to fill the gap between the infinities appearing in the QFT 
renormalized results and the finite values obtained in the literature 
with other procedures.



\newpage

\section{Introduction}

The question, most beautifully phrased by Eugene Wigner as that of
{\it the unreasonable effectiveness of mathematics in the natural
sciences} \cite{ew1} is an old and intriguing one. It certainly
goes back to Pythagoras and his school ({\it ``all things are
numbers''}), even probably to the sumerians, and maybe to more
ancient cultures, which left no trace. In more recent times, I.
Kant and A. Einstein, among others, contributed to this idea with
extremely beautiful and profound reflections. Also, {\it
mathematical simplicity, and beauty,} have remained for many years, and to many
important theoretical physicists, crucial ingredients when having
to choose among different plausible possibilities.

A crude example of {\it unreasonable effectiveness} is provided by
the regularization procedures in quantum field theory based upon
analytic continuation in the complex plane (dimensional,
heat-kernel, zeta-function regularization, and the like). That one
obtains a physical, experimentally measurable, and extremely
precise result after these weird mathematical manipulations is, if
not unreasonable, certainly very mysterious. No wonder that for
more one highly honorable physicist, the taming of the infinities
that plague the calculations of modern quantum physics and the
ensuing regularization/renormalization issue, were always
unacceptable, illegal practices. Those methods are presently full
justified and blessed with Nobel Prizes, but more because of the
many and very precise experimental checkouts (the {\it
effectiveness}) than for their intrinsic {\it reasonableness}.

As this note is meant not just for specialists, a simple example may be
clarifying. Consider the calculation of the zero point energy (vacuum to
vacuum transition, also called Casimir energy \cite{Casimir}) 
corresponding to a quantum
operator, $H$, with eigenvalues $\lambda_n$,
\begin{eqnarray}
E_0 =   < 0 | H |  0 > = \frac{1}{2} \sum_n \lambda_n, \label{e1}
\end{eqnarray}
where the sum over $n$ is a sum over the whole spectrum (which
will involve, in general, several continuum and several discrete
indices.) The last appear typically when compactifying the space
coordinates (much in the same way as time compactification gives
rise to finite-temperature field theory). In turns out, in
fact, that only in very special cases will this sum yield a
convergent series. Generically one has to deal with a divergent
sum, which is then regularized by different appropriate means. The
zeta-function method \cite{zb1} ---which stands on solid and
flourishing mathematical grounds \cite{zbas1}--- will interprete
this sum as being the value obtained for the zeta function of the
operator $H$, namely
\begin{eqnarray}
\zeta_H (s) =  \sum_n \lambda_n^{-s}, \label{z1}
\end{eqnarray}
at the point $s=-1$ (we set $\hbar = c =1$). It turns out 
generically that $\zeta_H (s)$ is well
defined as an absolutely convergent series for Re $s > a_0$ ($a_0$ is a
certain abscissa of convergence), and that it can be analytically continued
(in a perfectly defined way) to the whole complex plane, with the
possible appearance of poles as only singularities. If, as it
often happens,  $\zeta_H (s)$ has no pole at $s=-1$ we are done;
if it hits a pole there, then further elaboration of the method is
necessary. That the mathematical result one gets through this
process of analytical continuation in the complex plane, does coincide
with the experimental result, constitutes the specific example of
{\it unreasonable effectiveness of mathematics} we are referring
to.

As one should expect, things do not turn out to be so simple
 and clear-cut in practice. Actually, one cannot assign a meaning to
the {\it absolute} value of the zero-point energy, and any
physically measurable effect is to be worked out as an energy
difference between two situations, such as a quantum  field in
curved space as compared with the same field in flat space, or a
field satisfying boundary conditions on some surface as compared
with the same field in the absence of the surface, etc. This
energy difference is the Casimir energy:
\begin{eqnarray}
E_C = E_0^{BC} - E_0 =  \frac{1}{2} \left( \mbox{tr } H^{BC} - \mbox{tr
} H \right). \label{ec1}
\end{eqnarray}
 
And it is here where the problem appears.
Imposing mathematical boundary conditions on physical quantum
fields turns out to be a highly non-trivial act. This was
discussed in much detail in a paper by Deutsch and Candelas
a quarter of a century ago \cite{dc79}. These authors quantized 
electromagnetic and scalar fields in the region near an 
arbitrary smooth boundary, and calculated the renormalized vacuum 
expectation value of the stress-energy tensor, to find that the 
energy density diverges as the boundary is approached. Therefore, 
regularization and renormalization did not seem to cure the 
problem with infinities in this case and an infinite {\it physical} 
energy was obtained if the mathematical boundary conditions were 
to be fulfilled. However, the
authors argued that, in nature,  surfaces have non-zero depth,
and its value could be taken as a handy (dimensional) cutoff in order
to regularize the infinities. This approach will be recovered later 
in the present paper. Just two years after Deutsch and Candelas' 
work, Kurt Symanzik carried out a rigorous analysis of quantum field 
theory in the presence of boundaries \cite{ks81}. Prescribing the value
of the quantum field on a boundary means using the Schr\"odinger
representation, and Symanzik was able to show rigorously that such
representation exists to all orders in the perturbative expansion.
He showed also that the field operator being diagonalized in a smooth
hypersurface differs from the usual renormalized one by a factor
that diverges logarithmically when the distance to the
hypersurface goes to zero. This requires a precise limiting
procedure, and also point splitting, to be applied. In any case,
the issue was proven to be perfectly meaningful within the domains
of renormalized quantum field theory. One should note that in this
case the boundary conditions and the hypersurfaces themselves were
always treated at the pure mathematical level (zero depth) by
using delta functions.

More recently, a new approach to the problem has been postulated
\cite{bj1} which uses elements of the two methods above. Boundary
conditions on a field, $\phi$, are enforced on a surface, $S$, by
introducing a scalar potential, $\sigma$, of Gaussian shape living
on the surface (and near it). In the limit when the Gaussian
becomes a delta function, the boundary conditions (Dirichlet, in this case) 
are enforced, in that the delta-shaped potential kills all the
modes of $\phi$ at the surface. For the rest, the quantum system
undergoes a full-fledged quantum field theory renormalization,
as in the case of Symanzik's approach. The results obtained
confirm those of \cite{dc79}, in the several models studied albeit
they do not seem to agree with those of \cite{ks81}.

Such results are also in clear contradiction with the ones quoted in
the usual textbooks and review articles dealing with the Casimir effect
\cite{cb1}, where no infinite energy density when approaching any of the
Casimir plates has been reported.


\section{A zeta-function approach}

Too often has it been argued that  sophisticated  regularization
methods, as for instance the zeta-function regularization
procedure, get rid of the infinities in
 a rather obscure way (through analytic continuation, as previously
described), so that, contrary to what happens when a cut off is
introduced, one cannot keep trace of the infinites, which are 
cleared up without control, leading sometimes to erroneous results.

There is no way to refute an statement of this kind rigorously, 
but it should be pointed out that in many ocasions (if not
always) the discrepancies between the result obtained by using the
zeta procedure and other, say cut-off like, approaches have been proven
to emerge from a {\it misuse} of zeta regularization, and {\it not} to stem
from the method itself. When employed properly, the correct results have
been recovered (for a good number of examples, see
\cite{zb1,zbas1,bp1,zb3}).

Take the most simple example of a scalar field in one dimension,
$\phi (x)$, with a boundary condition (BC) of Dirichlet type
imposed at a point (say $x=0$), e.g. $\phi (0)=0$. We would like
to calculate the Casimir energy corresponding to this
configuration, that is, the difference between the zero point
energy corresponding to this field when the boundary condition is
enforced, and the zero point energy in the absence of any boundary
condition at all. Taken at face value, both energies are infinite.
The regularized difference may still be infinite when the BC point
is approached (this is the result in \cite{bj1}) or might turn out to
be finite (even zero, which is the result given in some standard
books on the subject).

Let us try to understand this discrepancy.
According to Eq. (\ref{e1}), we have to add up all energy modes.
For the mode  with energy $\omega$, the field equation reduces to:
\begin{eqnarray}
-\phi''(x)+m^{2}\phi (x)=\omega^{2}\phi(x). \label{fe1}
\end{eqnarray}
In the absence of a BC, the solutions to the field equation can be
labeled by $k = + \sqrt{\omega^2 - m^2} > 0$, as
\begin{eqnarray}
\phi_k (x)= A e^{ikx} + B e^{-ikx},
\end{eqnarray}
with $A,B$ arbitrary complex (for the general complex), or as
\begin{eqnarray}
\phi_k (x)= a \sin (kx) + b \cos (kx),
\end{eqnarray}
with $a,b$ arbitrary real (for the general real solution). Now, when the
mathematical BC of Dirichlet type, $\phi (0)=0$, is imposed, this
does not influence at all the eigenvalues, $k$, which remain
exactly the same (as stressed in the literature). However, the {\it number}
of solutions corresponding to each eigenvalue is reduced by one half to:
\begin{eqnarray}
\phi_k^{(D)} (x)= A (e^{ikx} -  e^{-ikx}),
\end{eqnarray}
with $A$ arbitrary complex (complex solution), and
\begin{eqnarray}
\phi_k^{(D)} (x)= a \sin (kx),
\end{eqnarray}
with $a$ arbitray real (real solution). In other words, the
energy spectrum (for omega) that we obtain in both cases is the same,
a continuous spectrum
\begin{eqnarray}
\omega = \sqrt{m^2 + k^2},
\end{eqnarray}
but the number of eigenstates corresponding to a given eigenvalue
is twice as big in the absence of the BC.\footnote{To understand this
point even better (by making recourse to what is learned in the
maths classes at high school), consider the fact that further, by
imposing Cauchy BC: $\phi(0)=0, \phi'(0)=0$, the eigenvalues
remain the same, but for any $k$ the family of eigenfunctions
shrinks to just the trivial one: $\phi_k(x)= 0, \forall k$
(the Cauchy problem is an initial value problem, which completely
determines the solution).}

Of course these considerations are elementary, but they seem to
have been put aside sometimes. They are crucial when trying to calculate
(or just to give sense to)  the Casimir energy density and force.
More to the point, just in the same way as the traces of the two 
matrices $ M_1= $ diag $(\alpha,\beta)$ and $ M_2= $ diag 
$(\alpha,\alpha,\beta,\beta)$
are not equal in spite of having ``the same spectrum $\alpha,\beta$,'' 
it also turns out that, in the problem under discussion, the traces of the  
Hamiltonian with and without the Diriclet BC imposed yield  different 
results, both of them divergent, namely
\begin{eqnarray}
 \mbox{tr } H = 2 \mbox{ tr } H^{BC} = 2 \int_0^\infty dk \, \sqrt{m^2 + k^2}.
\label{ec2}
\end{eqnarray}
By using the zeta function, we define
\begin{eqnarray}
\zeta^{BC} (s) :=  \int_0^\infty d\kappa \, (\nu^2 +
\kappa^2)^{-s}, \quad \nu := \frac{m}{\mu}, \label{zf2}
\end{eqnarray}
with $\mu$ a regularization parameter with dimensions of
mass.\footnote{Always necessary in zeta regularization, since the
complex powers of the spectrum of a (pseudo--) differential operator
can only be defined, physically, if the operator is rendered
dimensionless, what is done by introducing this parameter. That is
also an important issue, which is sometimes overlooked.} We get
\begin{eqnarray}
\zeta_{BC} (s) = \frac{\sqrt{\pi} \ \Gamma (s - 1/2)}{2 \, \Gamma
(s)} \, (\nu^2)^{1/2-s}, \label{zf3}
\end{eqnarray}
and consequently,
\begin{eqnarray}
 \mbox{tr } H^{BC}&=& \frac{1}{2} \zeta_{BC} (s=-1/2) \nonumber \\ 
& =&  \frac{m^2}{4 \,
 \sqrt{\pi}}
\left. \left[ \frac{1}{s+1/2} + 1 - \gamma - \log
\frac{m^2}{\mu^2} - \Psi (-1/2) + {\cal O} ( s+1/2 ) \right]
\right|_{s=-1/2}. \label{zf4}
\end{eqnarray}
As is obvious, this divergence is {\it not cured} when taking the
difference of the two traces, Eq. (\ref{ec1}), in order to obtain
the Casimir energy:
\begin{eqnarray}
E_C/\mu = E_0^{BC}/\mu - E_0/\mu = -   E_0^{BC}/\mu = \frac{\Gamma
(-1) m^2}{8\mu^2}. \label{ec3}
\end{eqnarray}
Wejust hit the pole of the zeta function, in this case.

How is this infinite to be interpreted? What is its origin? Just by
making recourse to the pure mathematical theory ({\it reine
Mathematik}), we already get a perfect description of what happens
and understand well where does  this
infinite energy\footnote{In mathematical terms, this infinite value
for the trace of the Hamiltonian operator.}
 come from. It clearly originates from the fact that imposing the boundary
condition has reduced to one-half the family of eigenfunctions
corresponding to any of the eigenvalues which constitute the
spectrum of the operator. And we can also advance that, since this
dramatic reduction of the family of eigenfunctions takes place
precisely at the point where the BC is imposed, the physical divergence
 (infinite energy) will  originate right there, and nowhere else.

While the analysis above cannot be taken as a substitute for the
actual modelization of Jaffe et al. \cite{bj1} ---where the BC is
explicitly enforced through the introduction of an auxiliary, localized
field, which probes what happens at the boundary in a much more precise
way--- it certainly shows that {\it pure mathematical
considerations}, which include the use of analytic continuation by
means of the zeta function, are in no way blind to the infinites
of the physical model and {\it do not} produce misleading results, when
the mathematics are used properly. And it is very remarkable to realize
how close the mathematical description of the appearance of an
infinite contribution is to the one provided by the physical realization 
\cite{bj1}.

\section{The case of two-point Dirichlet boundary conditions}

A similar analysis can be done for the case of a two-point
Dirichlet BC:
\begin{eqnarray}
\phi (a) =0, \qquad \phi (-a) =0. \label{2bc1}
\end{eqnarray}
Straightforward algebra shows, in this situation, that the eigenvalues
$k$ are quantized, as $k=\dot{\pi} /(2 a)$, so that:
\begin{eqnarray}
\omega_\ell = \sqrt{m^2+ \frac{\ell^2 \pi^2}{4 a^2}}, \quad \ell =
0, 1,  2, \ldots \label{2bc2}
\end{eqnarray}
The family of eigenfunctions corresponding to a given eigenvalue,
$\omega_\ell$, is of continuous dimension 1, exactly as in the
former case of a one-point Dirichlet BC, namely,
\begin{eqnarray}
\phi_\ell (x)= b \sin \left( \frac{\ell\pi}{2a} (x-a) \right),
\label{2bc3}
\end{eqnarray}
where $b$ is an arbitrary, real parameter.\footnote{The
contribution of the zero-mode ($\ell =0$) is controverted, but we
are not going to discuss this issue here (see e.g. \cite{et1} and 
references therein.)} To repeat,
the act of imposing Dirichlet BC on two points has the effect
of discretizing the spectrum (as is well known) but there is no
further shrinking in the number of eigenfunctions corresponding to
a given (now discrete) eigenvalue.

The calculation of the Casimir energy, by means of the zeta function, 
proceeds in this case as follows \cite{zb1,zbas1,bp1,zb3}. To begin 
with, it may be interesting to recall that the zeta-`measure' of 
the continuum
equals  twice the zeta-`measure' of the discrete. In fact, just
consider the following regularizations:
\begin{eqnarray}
\sum_{n= 1}^\infty \mu = \left. \mu  \sum_{n= 1}^\infty n^{-s}
\right|_{s=0} =
 \mu \zeta_R (0) = -\frac{ \mu }{2},
\end{eqnarray}
and
\begin{eqnarray}
\int_\mu^\infty d k = \left. \int_0^\infty d k \, (k + \mu)^{-s}
\right|_{s=0} = \left.\frac{\mu^{1-s}}{s-1}\right|_{s=0} = -  \mu.
\end{eqnarray}
The result is, as advanced, that the $\zeta$ `measure' of the
discrete is half that of the continuum.

The trace of the Hamiltonian corresponding to the quantum system with
the BC imposed, in the massive case, is obtained by means of the zeta 
function
\begin{eqnarray}
 \hspace*{-6mm}\zeta^{BC} (s) &:=&  \sum_{\ell =1}^\infty \left(
\frac{m^2}{\mu^2} + \frac{\pi^2 \ell^2}{4 \mu^2 a^2} \right)^{-s}
 \nonumber \\ && \hspace*{-17mm} =
 \left( \frac{ \mu}{m} \right)^{2s} \left[ -\frac{1}{2}
 + \frac{\Gamma (s-1/2)}{\Gamma (s)}\, \frac{ a
m}{\sqrt{\pi}} + \frac{2
\pi^s}{\Gamma (s)}\left( \frac{2 a m}{\pi}\right)^{1/2+s}
\sum_{n=1}^\infty n^{s-1/2} K_{s-1/2} (4anm) \right],
\end{eqnarray}
being  $K_\nu$ a modified Bessel function of the third kind (or MacDonald's 
function). Thus, for the zero point energy of the system  with two-point
Dirichlet BC, we get
\begin{eqnarray}
 \mbox{tr } H^{BC}/\mu = \frac{1}{2} \zeta_{BC} (s=-1/2) =-\frac{\Gamma (-1)
  m^2}{8 \mu^2}
 -\frac{m}{2\pi\mu} \ \sum_{n=1}^\infty \frac{1}{n} K_1 (2 \pi nm/\mu),
\end{eqnarray}
where $\mu$ is, in this case, $\mu:= \pi/(2a)$ ($a$ fixes the
mass scale in a natural way here). As in the previous example, we finally 
obtain an infinite value for the Casimir energy, namely
\begin{eqnarray}
E_C/\mu = E_0^{BC}/\mu - E_0/\mu  =  \frac{\Gamma
(-1) m^2}{8\mu^2} -\frac{m }{2\pi\mu} \ \sum_{n=1}^\infty \frac{1}{n}
K_1 (2 \pi nm/\mu) . \label{ec4}
\end{eqnarray}

It is, therefore, not true that regularization methods using
analytical continuation (and, in particular, the zeta approach)
are unable to see the infinite energy that is generated on the
boundary-conditions surface \cite{dc79,ks81,bj1} (see Eq. (\ref{e11})
later). The reason is still the
same as in the previous example: imposing a two-point Dirichlet
BC amounts again to halving the family of eigenfunctions
which correspond to any given eigenvalue (all are discrete, in the
present case, but this makes no difference). In physical terms,
that means having to apply an infinite amount of energy on the BC
sites, in order to enforce the BC. In absolute analogy, from the 
mathematical
viewpoint, halving the family of eigenfunctions immediately results
in the appearance of an infinite contribution, under the form of a
pole of the zeta function.

The reason why these infinities (the one here and that in the previous
section) do not usually show up in the literature on the
Casimir effect is probably because the textbooks on the subject 
focus towards
the calculation of the Casimir {\it force}, which is obtained by
taking minus the derivative of the energy 
with respect to the plate (or point) separation (here w.r.t.~$2a$). Since the
infinite terms do not depend on $a$, they do not contribute to the
force (as is recognized explicitly in \cite{bj1}). However, some
erroneous statements have indeed appeared in the above mentioned
classical references, stemming from the lack of recognition of the
catastrophical implications of the act of halving the number of
eigenfunctions, when imposing the BC. The persistence of the eigenvalues 
of the spectrum was probably misleading.  We hope to have clarified this
issue here.

\section{How to deal with the infinities}

Here, the infinite contributions have shown up at the regularization 
level, but
a more careful study \cite{bj1} was able to prove that they
do not disappear even after renormalizing in a proper way. The
important question is now: are these infinities {\it physical}? Will they
be {\it observed} as a manifestation of a very large energy pressure when
approaching the BC surface in a lab experiment? No doubt such questions 
will be best answered in that way, e.g. experimentally. 
If, on the contrary, that sort of large pressures fails to manifest itself,
this might be a clear indication of the need for an
additional regularization prescription. In principle, this seems to be
forbidden by standard renormalization theory, since the procedure has been
already carried out to the very end: there remains no additional physical 
quantity which could possibly absorb the divergences (see \cite{bj1}).

In any case, there are circumstances ---both  in physics and in
mathematics--- where  certain  `non-orthodox'
regularization methods have been employed with promising success. In
particular, Hadamard regularization in higher-post-Newtonian
general relativity \cite{had1} and also in recent variants of axiomatic and 
constructive quantum field theory \cite{had2}.  Among
mathematicians, Hadamard regularization is nowadays a rather  standard
technique in order to deal with singular differential and 
integral equations with boundary conditions, both
analytically and numerically (for a sample of references see
\cite{had3}). Indeed,
Hadamard regularization is a well-established procedure in order to 
give sense to infinite integrals. It is not to be found in the classical 
books on infinite calculus by Hardy or Knopp; it was  L. Schwartz 
\cite{sch1} who popularized it, rescuing Hadamard's original papers.
Nowadays, Hadamard convergence is one of the cornerstones in the
rigorous formulation of QFT through micro-localization, which on its turn is 
considered by specialists to be the most important step towards the 
understanding of linear PDEs since the invention of distributions (for a 
beautiful, updated treatment of Hadamard's regularization see \cite{melr1}).

Let us briefly recall this formulation. Consider a function, $g(x)$, 
expandable as
\begin{eqnarray}
 g(x)=\sum_{j=1}^k \frac{a_j}{(x-a)^{\lambda_j}} +h(x),
\end{eqnarray}
with $\lambda_j$  complex in general and $h(x)$ a regular function.
Then, it is immediate that
\begin{eqnarray}
 \int_{a+\epsilon}^b dx \, g(x) =P(1/\epsilon) + H(\epsilon),
\end{eqnarray}
being $P$ a polynomial and $H(0)$ finite. If the 
$\lambda_j\notin \mbox{\bf N}$, then one defines the Hadamard regularized 
integral as
\begin{eqnarray}
=\hspace*{-4.3mm}\int_{a}^b  dx \, g(x)  \ := \ \int_{a}^b h(x) \ dx
-\sum_{j=1}^k \frac{a_j}{\lambda_j-1} (b-a)^{1- \lambda_j}.
\end{eqnarray}
 Alternatively, one may define, for $ \alpha \notin \mbox{\bf N}, \
p<\alpha < p+1$, and $f^{(p+1)} \in C_{[-1,1]}$,
\begin{eqnarray}
\hspace*{-10.5mm} K^\alpha f : = \frac{1}{\Gamma (-\alpha)}
=\hspace*{-4.5mm}\int_{-1}^1   dt\ \frac{f(t)}{(1-t)^{\alpha +1}}, 
\end{eqnarray}
to obtain, after some steps,
\begin{eqnarray}
 \hspace*{-11mm} K^\alpha f = \sum_{j=0}^p \frac{f^{(j)}
 (-1)}{\Gamma (j+1-\alpha) \, 2^{\alpha -j}} +\frac{1}{\Gamma (p+1-\alpha)} \,
-\hspace*{-4.3mm}\int_{-1}^1  (1-t)^{p-\alpha}
f^{(p+1)}(t),
\end{eqnarray}
where the last integral is at worst improper (Cauchy's principal part). 
If $\lambda_1=1$, then the result is $ a_1 \ln (b-a)$, instead. 
If $\lambda_1 =p \in \mbox{\bf N}$,  calling 
\begin{eqnarray}
  H_p(f;x) \, := \ \
=\hspace*{-4.5mm}\int_{-1}^1  dt\ \frac{f(t)}{(t-x)^{p+1}} , \qquad
\qquad  { |x| < 1},
\end{eqnarray}
we get
\begin{eqnarray}
  H_p(f;x)
=\int_{-1}^1 \left[ f(t)-\sum_{j=0}^p \frac{f^{(j)}(x)}{j!}
(t-x)^j\right] \frac{dt}{(t-x)^{p+1}} \ + \ \frac{f^{(j)}(x)}{j!}
=\hspace*{-4.5mm}\int_{-1}^1 \frac{dt}{(t-x)^{p+1-j}}, 
\end{eqnarray}
where the first term is regular and the second one can be easily reduced to
\begin{eqnarray}
\frac{1}{(p-j)!}\ {\frac{d^{p-j}}{dx^{p-j}}}
-\hspace*{-4.3mm}\int_{-1}^1 \frac{dt}{t-x},  
\end{eqnarray}
being the last integral, as before, a Cauchy PP.

 An alternative form of Hadamard's regularization, which is more fashionable 
for physical applications (as is apparent from the expression itself) is the 
following \cite{had1}. For the case of two singularities,  at 
$\vec{x}_1, \ \vec{x}_2$, after excising from space two little balls 
around them,  $\mbox{\bf R}^3
\backslash \left( B_{r_1}(\vec{x}_1) \cup B_{r_2}(\vec{x}_2)\right)$,
with $B_{r_1}(\vec{x}_1) \cap B_{r_2}(\vec{x}_2) = \emptyset$, one defines
the regularized integral as being the limit
\begin{eqnarray}
 =\hspace*{-4.4mm}\int \,  d^3x \ F(\vec{x}) \ := \ 
\mbox{FP}_{\alpha,\beta
\to 0} \int d^3x \, \left(\frac{r_1}{s_1} \right)^\alpha
\left(\frac{r_2}{s_2} \right)^\beta F(\vec{x}),
\end{eqnarray}
where $s_1$ and $s_2$ are two (dimensionfull) regularization parameters 
\cite{had1}. This is the version that will be employed in what follows.

\section{Hadamard regularization of the Casimir effect}

We now use Hadamard's regularization as an additional tool in order to 
make sense of the infinite expressions encountered in the boundary value 
problems considered 
before. As it turns out from a detailed analysis of the results in 
\cite{bj1} (which we shall not repeat here, for conciseness), the basic 
integrals which produce infinities, in the one-dimensional and two-dimensional 
cases there considered, are the following. 

In one dimension, with Dirichlet BC imposed at one ($x=0$) and two
($x=\pm a$) points, 
respectively, by means of a delta-background of strength $\lambda$ (see 
\cite{bj1}), one encounters the two divergent integrals:
\begin{eqnarray}
 E_{1}(\lambda,m) & = &\frac{1}{2\pi}\int_{m}^{\infty}
\frac{dt}{\sqrt{t^2-m^2}}\,
\left[t\log\left(1+\frac{\lambda}{2t}\right)-\frac{\lambda}{2}\right], 
 \label{d11} \\
 E_2(a,\lambda,m) &=& \frac1{2\pi} \int_m^\infty \frac{dt}{\sqrt{t^2-m^2}}\,
           \left\{ t\log \left[1+\frac\lambda{t} + \frac{\lambda^2}{4t^2}
           \left(1-e^{-4at}\right)\right] - \lambda\right\}.
           \label{d12}
\end{eqnarray}
Using Hadamard's regularization, as described before, we obtain for the
first one, Eq. (\ref{d11}),
\begin{eqnarray}
 E_{1}(m) = \left. \frac{\lambda}{4\pi} \left(1-\ln \frac{\lambda}{m} \right) 
\right|_{\lambda \to \infty}  + \
 =\hspace*{-4.5mm}\int \ \, , \label{e11}
\end{eqnarray}
where the first term is the singular part when the limit $\lambda \to \infty$ 
is taken, and the second ---which is Hadamard's finite part--- yields in this 
case
\begin{eqnarray}
  =\hspace*{-4.5mm}\int \ = \, -\frac{m}{4}.
\end{eqnarray}
Such result is coinciding with  the classical
one (0, for $m=0$). Note in particular, that the further 
$ \ln m $ divergence 
as $ m \to \infty $ is hidden in the  $\lambda-$divergent part, and 
that behavior does explain why the classical results which are obtained 
using hard Dirichlet BC ---what corresponds as we just prove here 
to the Hadamard's regularized part--- cannot see anything of it.

In the case of a two-point boundary at $x=\pm a$ (separation $2a$),  
Eq. (\ref{d12}), we get a similar Eq. (\ref{e11}) but now the regularized 
integral is as follows. For the massless case, we obtain
\begin{eqnarray}
  =\hspace*{-4.5mm}\int \ = \, -\frac{\pi}{48 a},
\end{eqnarray}
which is  the regularized result to be found in the classical books. In the 
massive case, $m\neq 0$, after some additional work the following fast 
convergent series turns up [cf. Eq. (\ref{ec4})]
\begin{eqnarray}
 =\hspace*{-4.5mm}\int \ = \,  -\frac{m}{2\pi} \sum_{k=1}^\infty
\frac{1}{k} K_1 (4akm).
\end{eqnarray}
Thus Eq. (\ref{e11}) yields strictly the same result (\ref{ec4}) that was 
already obtained by imposing the Dirichlet BC {\it ab initio}. What has now
been {\it gained} is a more clear identification of the singular part, in 
terms of the strength of the delta potential at the boundary. This will be the 
general conclusion, common to all the other cases here considered.

Correspondingly, for the Casimir force we obtain the finite 
values\footnote{Note that the force $F(a)$ is here given by minus the 
derivative of the total energy $E(a)$  w.r.t.~$2a$, since this is the 
distance between the two Dirichlet points ({\it not} $a$.)}
\begin{eqnarray}
 F_2(a) = -\frac{\pi}{96 a^2},
\end{eqnarray}
in the massless case, and in the massive one
\begin{eqnarray}
  F_2(a,m)  =  -\frac{m^2}{\pi} \sum_{k=1}^\infty \left[
K_0 (4akm) +\frac{1}{4akm} K_1 (4akm) \right].
\end{eqnarray}
Those expressions coincide with the ones derived in the above mentioned 
textbooks on the Casimir effect, and reproduced before by using the zeta
function method (just take minus the derivative of Eq. (\ref{e11}) 
w.r.t.~$2a$).

The two dimensional case turns out to be more singular  \cite{bj1} 
---in part just for dimensional reasons--- and requires additional 
wishful thinking
in order to deal with the circular delta function sitting on the 
circumference where the Dirichlet BC are imposed. Here one encounters the 
basic singular integral, for the term contributing to the second Born 
approximation (we use the same notation as in \cite{bj1}),
\begin{eqnarray}
\tilde{\sigma} (p)=\int_0^\infty dr \, r  J_0(pr) \sigma (r), \qquad
\sigma (r) = b \lambda \exp \left[ -\frac{(r-a)^2}{2\omega^2}
\right] ,
\end{eqnarray}
 with $J_0$ a Bessel function of the first kind, and
\begin{eqnarray}
 \int_0^\infty  dr\, \sigma (r) =
\lambda, \qquad   \sigma (r) \ 
\stackrel{\omega \to 0}{\longrightarrow} 
\ \lambda \, \delta (r-a).
\end{eqnarray}
Hadamard's regularization yields, in this case (the $\tau$'s replacing 
the $\sigma$'s in the regularized version),
\begin{eqnarray}
 \tau (r,p)= c \lambda (rp+1)^{-\omega /2} \exp
\left[ -\frac{(r-a)^2}{2\omega^2} \right] \ \stackrel{\omega \to
0}{\longrightarrow} \ \lambda \, \delta (r-a),
\end{eqnarray}
with $p$ a (dimensionfull) regularization parameter, 
being the constant $c$ given by
\begin{eqnarray}
 c^{-1}=\int_0^\infty dr \, r^{-\omega}\exp \left[
-\frac{(r-a)^2}{2\omega^2} \right],
\end{eqnarray}
which exists and is perfectly finite; in particular,
$ c^{-1} (\omega =.1, a=1) =.25$. Then,
\begin{eqnarray}
\tilde{\tau} (p) = 2 \pi \int_0^\infty dr \, r \, J_0(pr) \,  \tau (r,p)
=2\pi \lambda a (ap+1)^{-\omega /2}  J_0(ap)
\end{eqnarray}
It turns out that for the Casimir energy we get in this case (notation as
in \cite{bj1})
\begin{eqnarray}
 \hspace*{-4mm} E^{(2)}_{\lambda^2} [\tau] &=&
\left. \frac{\lambda^2a^2}{8}\int_0^\infty dp \,
(ap+1)^{-\omega} J_0(ap)^2 \arctan (p/2m)  \right|_{\omega \to 0} \nonumber \\
&=& \frac{\lambda^2a^2}{8} \left\{
 \frac{1}{2\omega} \ +  \frac{\gamma + 3 \ln 2}{2a}
+4m \left[ \gamma - \frac{2}{\sqrt{\pi}} \left[ 1- \ln
(am) \right] \, h(4a^2m^2) \right]\right\},
\end{eqnarray}
where 
$ h(z):= {_2F_3} \left( (1/2,1/2);(1,1,3/2);z \right)$ and $\gamma$ is 
the Euler-Mascheroni constant; in particular, for instance 
$h(1)=1.186711$, what is quite a nice value.
Recall also that  $ \omega$ is the  {\it width} of the Gaussian  
 $ \delta$, which is
the very {\it physical} parameter considered in \cite{dc79}. 
When this width tends to zero
an infinite energy appears (the width controls the formation of the pole).
The rest of the result is the Hadamard regularization of the 
integral, e.g.\footnote{It should be pointed out that the 
computational program Mathematica \cite{math1}
 directly assigns the Hadamard regularized 
value to particular cases of integrals of this kind; but it does so without 
any hint on what is going on. This has often confused more one user, who 
fails to understand how it comes that an infinite integral gets a finite
value out of nothing.}
\begin{eqnarray}
=\hspace*{-4.5mm}\int_0^\infty   dp\, J_0(ap)^2 \arctan (p/2m).
\end{eqnarray}
Again, the finite part reverts to the results obtained in the literature with
Dirichlet BC {\it ab initio}.

To summarize, it has been here proven ---in some particular but 
rather non-trivial and representative examples--- that the finite results 
derived through the use of Hadamard's regularization exactly 
coincide with the values obtained using the more classical,
less full-fledged methods to be found in the literature on the Casimir
effect. Moreover, Hadamard's prescription is able to separate and identify 
the singularities as physically meaningful cut-offs. 
Although the validity of this additional 
regularization is at present questionable, the fact that it bridges the
two approaches is already remarkable, maybe again a manifestation of the 
unreasonable effectiveness of mathematics.

\medskip

\noindent{\bf Acknowledgments}

The author is indebted to the members of the Mathematics Department, MIT, 
where this work was completed, and specially to Dan Freedman, for warm 
hospitality. This investigation has been supported by DGICYT (Spain), 
project BFM2000-0810 and by CIRIT (Generalitat de Catalunya), grants 
2002BEAI400019 and 2001SGR-00427.

\vspace{1mm}


\end{document}